\documentclass[aps,showpacs,superscriptaddress,longbibliography]{revtex4-2}

\usepackage{amsfonts}
\usepackage{amssymb}
\usepackage{amsmath}
\usepackage{graphicx,color}
\usepackage{bm}
\usepackage{ulem}
\usepackage{import}
\usepackage{lineno}
\usepackage{slashed}
\usepackage{braket}
\usepackage{dcolumn}% Align table columns on decimal point

\newcommand{\red}[1]{\textcolor{black}{#1}}

\newcommand{\be}{\begin{equation}}
\newcommand{\ee}{\end{equation}}
\newcommand{\bea}{\begin{eqnarray}}
\newcommand{\eea}{\end{eqnarray}}
\newcommand{\bref}[1]{(\ref{#1})}

\begin{document}

\title{Right-handed weak\red{-}currents in neutrinoless $\beta \beta $ decays and ton-scale $\beta\beta$ detectors}  

\author{Hiroyasu Ejiri}
\email{ejiri@rcnp.osaka-u.ac.jp}
\author{Takeshi Fukuyama}
\email{fukuyama@rcnp.osaka-u.ac.jp}
\author{Toru Sato}
\email{tsato@rcnp.osaka-u.ac.jp}
 \affiliation{Research Center for Nuclear Physics, Osaka University, Osaka 567-0047, Japan
 }%

\date{\today}% It is always \today, today,
\begin{abstract}

 Right handed weak-currents (RHCs) in the left-right (L-R) symmetric model for neutrinoless double beta decays (DBDs) of both the $0^+~\to~0^+$ and $0^+~\to~2^+$ transitions are discussed from both theoretical and experimental view points.
 \red {
   $<\lambda>$ and $<\eta>$-terms are related by
    $<\lambda>/<\eta> \approx \tan\beta$, which is constrained in the regions of $1-60$ for SUSY grand unified theories (GUTs) and of $1-165$ for non-SUSY GUTs.
%   \sout{New theoretical aspects of the RHCs are discussed on the basis of BSM (beyond the standard model) and the order of the magnitude of $<\lambda>/<\eta>$ is related to the tan$\beta$, which is constrained in the regions of $1-60$ for SUSY GUT and of $1-165$ for non-SUSY GUT.}
   The  enhancement mechanisms of the $<\eta>$ term over the $<\lambda>$ term in the $0^+$ transition are shown, and the $\Delta$ isobar contribution to the NME for the transition to the 2$^+$ state is found to be of the order of $20\%$ of the NME with the quenched weak coupling.  The new and interesting RHC regions of 
 $<\lambda>\approx 5\times10^{-8}$ and $<\eta>\approx 1.5\times 10^{-10}$ are shown to be exclusively explored by measuring  both the $\beta\beta $ and $\gamma$ rays associated with the ground and excited DBDs by means of the ton-scale DBD detectors for the IH (inverted hierarchy) $\nu$-masses. The actual RHCs to be studied depend on the RHC NMEs. }

\end{abstract}

\maketitle

\section{Introduction} ~~
Neutrinoless double beta decay (DBD)
is a realistic probe for studying the neutrino ($\nu$) nature (Majorana or Dirac), the absolute $\nu$-mass scale,  the right-handed weak current (RHC) and others, which are  beyond the standard model (BSM), as discussed in reviews \cite{doi85,eji05,avi08,eji10,ver12} 
and references therein. 

In the present work we study the effects of the RHC in the $0\nu$ DBDs to the ground and excited states, where $<\lambda>$- and $<\eta>$-terms  in addition to $<m>$-term  appear (Eq. \bref{LambdaEta}). Based on the neutrino oscillation experiments, the effective Majorana neutrino-mass $<m>$ is around 15-45 meV and 2-5 meV for the inverted mass hierarchy (IH) and the normal mass hierarchy (NH) cases, respectively. Then recent experimental works are concentrated mostly on the light-mass mechanism by measuring  the ground-state 0$^+\rightarrow$0$^+$ DBDs \cite{ago23,ume24}.

The DBD transition rate depends on the nuclear matrix element (NME). The NME, which reflects the complex nuclear correlations in the DBD nucleus, is crucial to extract the $\nu$-mass and the contributions of RHCs. The NME, however, depends on the models and the nuclear parameters used there \cite{suh98,fae98,suh12, eng17,eji19a}. One critical parameter is the effective axial-vector weak coupling
$g_{\rm A}^{eff}$, which is  much quenched in a nucleus due to the various kinds of the nucleonic and non-nucleonic correlations \cite{eji00,ste15,suh17,eji19}. 

Recent high-sensitivity experiments have excluded the  larger $\nu$-mass region of  $<m>\ \ge$\ 100-200 meV, depending on the NME.  Then they are going to build higher-sensitivity 
and larger-scale DBD detectors to search for the neutrino mass in the IH neutrino-mass region around $<m>\approx$  30 meV. Since the minimum neutrino-mass to be measured is proportional to ($B/N)^{1/4}$ with $N$ being the total amounts of the DBD isotopes used in the detector and $B$ being the backgrounds, one needs ton-scale ($N\approx$\ tons) detectors to measure the IH neutrino-mass and even multi-10 ton scale DBD detectors for the NH neutrino mass. Accordingly, one might anticipate that the ton-scale DBD detectors under construction may not be quite effective if the neutrino mass spectrum is of NH without RHC.  

>From the theoretical view point of grand unified theories (GUTs), our results prefer the NH to the IH: that is, inputting the observed lepton masses and the PMNS angles into the model, we compare the outputs of quark masses and CKM matrices in the model with the observed data, and obtained $\chi^2\leq 1$ for the NH case and $\chi^2>200$ for the IH case \cite{FIM}.
On the other hand, recent measurements of baryonic acoustic oscillations of cosmic microwave background constrain \cite{DESI, Planck, ACT} $\sum_i m_{\nu i}<0.072 ~~\mbox{eV at}~~95\%~\mbox{C.L.}$
This bound disfavors IH at $3\sigma$ \cite{Smirnov2}. If this is the case, the future ton-scale DBD detectors may find no $\nu$-mass signals but the RHC ones \cite{li21}. 

\red {DBD experiments, so far, have been made mostly for the ground $0^+$ state with the large phase space factors. In fact, the DBD phase space and thus the DBD rate for the excited 2$^+$ state are much smaller than those for the ground state, and in any way the $\nu$-mass term does not contribute to the excited-state transition as discussed later. Here we note that the excited-state DBD is characterized by the $\gamma$-ray from the excited state.}

\red {The present paper aims to report for the first time the new theoretical aspects of the RHCs and the new  RHC regions to be studied by measuring  both DBDs to the  ground and  excited states with the ton-scale DBD detectors for both the $\beta\beta$ and $\gamma$ rays.}
The RHCs to be studied are the $<\lambda>$ and $<\eta>$ in the region of the orders of $10^{-8}$ and $10^{-10}$, respectively. Then the DBD signal in the $0^+$ ground state  is due to either $<\lambda>$ or $<\eta>$, or both of them, while that in the $2^+$ one is exclusively due to $<\lambda >$, as will be discussed in Sec. 3 and Sec. 4. 
From these experiments \red {one obtains the values for the ratio of $<\lambda>/<\eta>$, which is related to tan$\beta$, and the parameters included in the RHC mechanisms in BSM physics}.

DBD detectors to be discussed in the present work are $\beta\beta-\gamma$ detectors to make it possible to separate the excited-state decay from the ground-state decay and to reduce considerably background contributions to both the ground- and excited-state DBD studies, 
as shown in early works with ELEGANT experiments \cite{eji87,eji96}.
Actually, DBD detectors used so far for high-sensitivity DBD experiments are mostly calorimetric detectors to measure the total DBD energy, but do not measure individual 2 $\beta$-ray tracks to identify the DBD processes, the mass, $\lambda$ and $\eta$ ones. 

This paper is organized as follows. The theoretical aspects of the RHC DBD are discussed briefly in section 2. The RHC NMEs for the 0$^+$ ground- and 2$^+$ excited-state transitions are discussed in section 3. Here we include brief discussions on the possible contribution of the $\Delta$ isobar to the RHC NMEs to the 2$^+$ excited state. Experimental sensitivities for measuring the $\lambda$ and $\eta$ RHCs and  experimental merits of measuring both $\beta \beta$ and $\gamma$ rays are explained in section 4. Finally concluding remarks on the future RCH DBD works are given in section 5. 

\section {Theoretical aspects of RHC DBDs}
\red{In discussing the theoretical models for $0\nu$ DBD the most important thing is how to make a neutrino mass.
There are two approaches; The first is the introduction of right-handed current, and the second the radiative generation of neutrino mass, both of which explain the reason why neutrino mass is much lighter than the other charged leptons and quarks. That is, the former is due to the seesaw mechanism and the latter to the loop factor $\left(\frac{1}{16\pi^2}\right)^n$ for n-loop
 \cite{FMU1}. Unfortunately, there are too many varieties of model options for the latter case. On the other hand, in the former case, RHC, has the firm root in GUT \cite{Fuku1} and is able to get the restrictions and correlations in BSM physics like Lepton Number or Flavour Violations etc. Therefore, in this paper we stand on the RHC scenario.}
We use the notations of \cite{FS} for the detailed applications of Left-Right (L-R) symmetric model.
The weak Hamiltonian is given by
 \begin{equation}
H_W=\frac{G_F\cos\theta_c}{\sqrt{2}}\left[j_{\rm L}^\mu \tilde{J}_{{\rm L}\mu}^\dagger+
   j_{\rm R}^\mu \tilde{J}_{{\rm R}\mu}^\dagger\right]+H.c.
\label{Fermi}
\end{equation}
Here $j_{\mu}$ ($J_{\mu}$) indicates the leptonic (hadronic) current, and the  L- and R-handed leptonic currents, $j_{{\rm L}\mu}$ and $j_{{\rm R}\mu}$, are given by
\begin{eqnarray}
\label{Leptonic1}
j_{{\rm L}\alpha}&=& \sum_{l=e,\mu,\tau}\overline{l(x)}\gamma_\alpha(1-\gamma_5)\nu_{lL}(x)\nonumber\\
&\equiv &\sum\overline{l(x)}\gamma_\alpha 2P_L\nu_{lL}(x),\\
j_{{\rm R}\alpha}&=& \sum_{l=e,\mu,\tau}\overline{l(x)}\gamma_\alpha(1+\gamma_5)N_{lR}(x)\nonumber\\
&\equiv& \sum\overline{l(x)}\gamma_\alpha 2P_RN_{lR}(x),
\label{Leptonic2}
\end{eqnarray}
where $\nu_{lL}(N_{lR})$ are L-handed (R-handed) weak eigenstates of the neutrinos, and
\bea
\label{Hadronic1}
  \tilde{J}_{\rm L}^\mu(\bm{x}) & = & J_{\rm L}^\mu(\bm{x}) + \kappa J_{\rm R}^\mu(\bm{x}),
  \\
  \tilde{J}_{\rm R}^\mu(\bm{x}) & = & \eta J_{\rm L}^\mu(\bm{x}) + \lambda J_{\rm R}^\mu(\bm{x}).
  \label{Hadronic2}
  \eea
The neutrino mass matrix is \cite{Minkowski, Yanagida, Gellmann, M_S}
\be
M_\nu=\left(
\begin{array}{cc}
M_L&M_D^T\\
M_D&M_R\\
\end{array}
\right)\approx
\left(
\begin{array}{cc}
0&M_D^T\\
M_D&M_R\\
\end{array}
\right).
\label{Mnu}
\ee
Thus we have the extended Fermi couplings \bref{Fermi}. In \bref{Leptonic1} and \bref{Leptonic2}, $\nu_{lL} (N_{lR})$ are L-handed (R-handed) weak eigenstates of the neutrinos. Using $3\times 3$ blocks $U,V,X,Y$, the mass eigenstates $\nu',~N'$ are given as
\be
\left(
\begin{array}{c}
\nu\\
(N_R)^c\\
\end{array}
\right)_L=\left(
\begin{array}{cc}
U&X\\
V^*&Y\\
\end{array}
\right)\left(
\begin{array}{c}
\nu'\\
N'\\
\end{array}
\right)_L
\equiv \mathcal{U}\left(
\begin{array}{c}
\nu'\\
N'\\
\end{array}
\right)_L.
\label{mixing}
\ee
That is,
\be
(\nu_L)_\alpha=U_{\alpha i}\nu_i'+X_{\alpha I}N_I',~~(N_R)^c_\alpha=V^*_{\alpha i}\nu_i'+Y_{\alpha I}N_I',
\label{U1}
\ee
where $\alpha$ ($i$) are the flavour (mass) eigenstates.
\be
 \mathcal{U}^TM_\nu\mathcal{U}=\left(
 \begin{array}{cc}
m_{light}&0_{3\times 3}\\
0_{3\times 3}&M_{heavy}
\end{array}
\right),
\label{U2}
\ee
\be
m_{light}=-M_D^TM_R^{-1}M_D.
\ee
>From GUT, if we adopted very naive order estimation neglecting flavour indices, $M_D$ is of order of the top quark mass, and $M_R$ is of order of $10^{14}$ GeV and the effect of RHC is not observable.
However, we can realize the TeV scale seesaw \cite{Smirnov} in the $SU(2)_L\times SU(2)_R\times U(1)_{B-L}$ 
model \cite{Mohapatra1}.
The transition rate (inverse of the half-life $T_{1/2}^J$)
for $0^+~\to~J^+ ~(J=0,2)$ is given as \cite{doi85, Simkovic1, Engel, Pantis2, Pantis, suhonen98} 
\begin{eqnarray}
\label{T1/2}
  \Gamma^{(J )}&\equiv &\frac{1}{T_{1/2}^J}=  C_{mm}^{(J)} (\frac{<m>}{m_e})^2+ C_{m\lambda}^{(J)}   \frac{<m>}{m_e}<\lambda> \nonumber\\
  &+& C_{m\eta}^{(J)}   \frac{<m>}{m_e}<\eta> + C_{\lambda\lambda}^{(J)} <\lambda>^2\\
  & + &C_{\eta\eta}^{(J)} <\eta>^2+ C_{\lambda\eta}^{(J)} <\lambda><\eta>. \nonumber
\end{eqnarray}
Here $C_{ab}^{(0)}$ includes the NME and the phase space integral. The other parts include the $\nu$-mass and the  BSM physics. The effective couplings $<\eta>$ and $<\lambda>$ are given as
\be
<m>=|\sum_iU_{ei}^2m_i|,~~<\lambda>=\lambda |\sum _j  U_{ej}V_{ej}|,~~<\eta>=\eta|\sum_j U_{ej}V_{ej}|.
\label{LambdaEta}
\ee

The constants $\lambda$ and $\eta$  in \bref{Fermi} are related to the mass eigenvalues of the weak bosons in the L and R-handed gauge sectors ($W_L,~W_R$) as follows \cite{Zhang}

\bea
\label{WL}
W_L^+&=&W_1^+\cos\zeta-W_2\sin\zeta e^{-i\alpha},\\
\label{WR}
W_R&=&W_1^+\sin\zeta e^{i\alpha}+W_2^+\cos\zeta,\\
\label{GF}
\frac{G_F}{\sqrt{2}}&=&\frac{g^2}{8}\cos^2\zeta\frac{M_{W1}^2\tan^2\zeta+M_{W2}^2}{M_{W1}^2M_{W2}^2}, \\
\label{lambda}
\lambda&= &\frac{M_{W1}^2+M_{W2}^2\tan^2\zeta}{M_{W1}^2\tan^2\zeta+M_{W2}^2},\\
\eta&= & -\frac{(M_{W2}^2-M_{W1}^2)\tan\zeta}{M_{W1}^2\tan^2\zeta+M_{W2}^2}.
\label{eta}
\eea
Here $M_{W1}$ and $M_{W2}$ are the masses of the mass eigenstates $W_1$ and $W_2$, respectively, and $\zeta$ is the mixing angle which relates the mass eigenstates and the gauge eigenstates.
We are considering L-R symmetric model. The gauge boson masses are \cite{Zhang}
\be
M_W^2=\left(
\begin{array}{cc}
\frac{1}{2}g^2(\kappa^2+\kappa'^2+2v_L^2)&-g^2\kappa\kappa'e^{-i\alpha}\\
-g^2\kappa\kappa'e^{i\alpha}&g^2v_R^2\\
\end{array}
\right).
\label{MW}
\ee
Here the the Yukawa coupling between quark doublets (and lepton doublets) and bi-doublet Higgs is given by
\be
\mathcal{L}_Y=\overline{Q}_{Li}(Y_{ij}\Phi+Y_{ij}'\tilde{\Phi})Q_{Rj}+H.c.
\label{LY}
\ee
with 
\be
<\Phi>=\left(
\begin{array}{cc}
\kappa & 0\\
0&\kappa'e^{i\alpha}\\
\end{array}
\right),~~\mbox{and}~~
\tilde{\Phi}=\tau_2\Phi^*\tau_2.
\label{Phi}
\ee
 Thus this model is the "minimal" L-R symmetric model. \red{In this mode, there are loop diagrams to generate $0\nu$ DBD other than the three tree diagrams \cite{FS, FMU1}. However it is subdominant compared with the above three tree diagrams.} 
We consider this model to clarify the essence of L-R symmetric model in DBD.
However, the quark mixing angle is almost diagonal and if we assume
\be
\kappa\gg \kappa',\ \ \mbox{and}\ \ \ Y_{ij}\gg Y_{ij}',
\label{approx0}
\ee
then they are related with more familiar quantities, 
\be
\kappa\approx v_u,~~\kappa'\approx v_d,~~\frac{1}{\sqrt{2}}(\kappa^2+\kappa'^2)=v_{ew}^2.
\label{approx}
\ee
Then the L-R weak boson mixing angle becomes
\be
\tan 2\zeta=-\frac{2\kappa\kappa'}{v_R^2}\approx \frac{2v_uv_d}{v_R^2}=-4\xi\left(\frac{M_{WL}}{M_{WR}}\right)^2
\label{zeta}
\ee
with
\be
\xi\equiv \kappa'/\kappa \approx v_d/v_u\equiv 1/\tan\beta
\label{xi}
\ee
\cite{Langacker} and
\be
M_{W2}=g_Rv_R\geq 3.7~\mbox{TeV}
\label{M2}
\ee
\cite{CMS}, where the symbol $\approx$ implies to adopt the assumption of \bref{approx0}. In the L-R symmetric model, we set $g_L=g_R$, which indicates further unification of at least rank five GUT, including SU(3) color. $\tan\beta$ is constrained from the fact that the Yukawa coupling is renormalizable up to the GUT scale, \red{
\be
1 \leq \tan\beta\leq \begin{cases}
60 & \mbox{SUSY case} \\
165 & \mbox{non-SUSY case}.
\end{cases}
\label{tanbeta}
\ee
That is, the upper (lower) limit comes from the renormalizability of both top and bottom Yukawa coupling at GUT (low energy) scale \cite{Shafi}.  For non-SUSY model, the upper limit can be as large as $165$.}   As will be shown in Sec. 4, we note that the RHC regions to be studied are $<\lambda> \approx 10^{-7}$ and $<\eta> \approx 10^{-10}$. If we adopt the approximation \bref{approx}, one may get 
\be
\lambda\approx \left(\frac{M_1}{M_2}\right)^2~ \mbox{and}~\eta\approx -\tan\zeta\approx \left(\frac{M_L^2}{M_R^2}\right)\frac{1}{\tan\beta}\approx \frac{\lambda}{\tan\beta}.
\ee

\section{RHC DBD NMEs for the ground $0^+$ and excited $2^+$ states}

\subsection{Transition amplitude of $0\nu$ DBD}

The transition amplitude  $I \rightarrow F + e_{p_1,s_1} + e_{p_2,s_2}$
 is given by the second order  perturbation of $H_W$ following Ref. \cite{Tomoda} as,
\begin{eqnarray}
  R_{0\nu} & = & \frac{4}{\sqrt{2}}(\frac{G\cos\theta_c}{\sqrt{2}})^2
  \sum_{\alpha,\beta,i} \int d\bm{x}d\bm{y}\int \frac{d\bm{k}}{(2\pi)^3}
  e^{-i\bm{k}\cdot \bm{r}}\frac{-1}{2\omega} \nonumber \\
    & &
\times [(\bar{e}_{\bm{p}_2,s_2}(\bm{y})
        \gamma^\mu P_\beta(\gamma_0 \omega - \bm{\gamma}\cdot\bm{k} + m_i)P_\alpha \gamma^\nu
  e^c_{\bm{p}_1,s_1}(\bm{x})    )
          (<F|\tilde{J}^{\nu \dagger}_\beta(\bm{y}) \frac{1}{E_I - (H_{st} + \omega + e_1)}  \tilde{J}^{\mu \dagger}_\beta(\bm{x})|I>)
          \nonumber \\
          & - &
 (\bar{e}_{\bm{p}_2,s_2}(\bm{y})
          \gamma^\mu P_\beta(\gamma_0 \omega + \bm{\gamma}\cdot\bm{k} - m_i)P_\alpha \gamma^\nu
          e^c_{\bm{p}_1,s_1}(\bm{x})    )
          (<F|\tilde{J}^{\nu \dagger}_\beta(\bm{y}) \frac{1}{E_I - (H_{st} + \omega + e_2)}  \tilde{J}^{\mu \dagger}_\beta(\bm{x})|I>)].
  \label{eq:r0nu}
\end{eqnarray}
Here the four momentum of the exchanged neutrino is $k^\mu = (\omega, \bm{k})$, $\bm{r} = \bm{x} - \bm{y}$
and $e_i$ is energy of electron, $\omega = \sqrt{\bm{k}^2 + m_\nu^2}$.
$|I>,|F>$ are state vector of the initial and the final hadron states
and $H_{st}$ is Hamiltonian of strong interaction.
It is noticed that at this stage hadron current $\tilde{J}^\mu_\alpha$
is given in terms of current quark  and initial and final state vector
can be nuclear states or any hadronic states.

To simplify the derivation, we assume that
the neutrino energy $\omega$ dominates
 the energy denominator $E_I - (H_{st} + \omega + e_i)$.
$R_{0\nu}$ is then expressed as
\begin{eqnarray}
  R_{0\nu} & = &
  \frac{i}{4\pi} \frac{4}{\sqrt{2}}
  (\frac{G\cos\theta_c}{\sqrt{2}})^2<F|{\cal M}|I>
\end{eqnarray}
with
\begin{eqnarray}
  {\cal M} & = &
\sum_{\alpha,\beta,i}  \int d\bm{x}d\bm{y} \frac{h(r)}{r^2}
  \bar{e}_{\bm{p}_2,s_2}(\bm{y})
  \slashed{\tilde{J}}^\dagger_\beta(\bm{y})  P_\beta (\bm{r}\cdot\bm{\gamma})  P_\alpha \slashed{\tilde{J}}^\dagger_\alpha(\bm{x})
  e^c_{\bm{p}_1,s_1}(\bm{x}).
  \label{eq:r0nu-1}
\end{eqnarray}
Here  $h(r)$ is given as
\begin{eqnarray}
  h(r) & = & -  \frac{4\pi}{(2\pi)^3}\ r \frac{d}{dr} \int d\bm{k}\frac{e^{-i\bm{k}\cdot\bm{r}}}{\omega^2} \sim \frac{1}{r}.
\end{eqnarray}

Now,
 the $\eta$ and $\lambda$ dependence of ${\cal M}$ are clearly given as
\begin{eqnarray}
  {\cal M} & = & \int d\bm{x}d\bm{y} \frac{h(r)}{r^2}
  \bar{e}_{\bm{p}_2,s_2}(\bm{y})  {\cal O}  e^c_{\bm{p}_1,s_1}(\bm{x}),
\end{eqnarray}
with
\begin{eqnarray}
  {\cal O} & = &
  <\eta> [
      (\slashed{V}(\bm{y})(\bm{r}\cdot\bm{\gamma})\slashed{V}(\bm{x})
      +    \slashed{A}(\bm{y})(\bm{r}\cdot\bm{\gamma})\slashed{A}(\bm{x}) )
      -  (  \slashed{V}(\bm{y})(\bm{r}\cdot\bm{\gamma})\slashed{A}(\bm{x})
      +    \slashed{A}(\bm{y})(\bm{r}\cdot\bm{\gamma})\slashed{V}(\bm{x}))]
      \nonumber \\
    & + &
      <\lambda> [
   ( \slashed{V}(\bm{y})   (\bm{r}\cdot\bm{\gamma}) \slashed{V}(\bm{x})
   - \slashed{A}(\bm{y})   (\bm{r}\cdot\bm{\gamma}) \slashed{A}(\bm{x}))
   -  ( \slashed{V}(\bm{y})(\bm{r}\cdot\bm{\gamma}) \slashed{A}(\bm{x})
   -   \slashed{A}(\bm{y})(\bm{r}\cdot\bm{\gamma}) \slashed{V}(\bm{x}))\gamma_5].
\end{eqnarray}
The difference between the  $<\eta>$- and $<\lambda>$-terms is transparent that
relative sign between  vector current $V-V$ and axial vector current $A-A$
and interference term of vector and axial vector term $V-A$ and $A-V$.
Furthermore, $\gamma_5$ of $<\lambda>$-term  modifies even-odd structure
of lepton Dirac matrix.
The formula can be used both for $0^+ \rightarrow 0^+$ and $0^+ \rightarrow 2^+$ nuclear DBD.
Since the angular momentum $L=1$ and parity $-1$
are carried by neutrino, %$\bm{r}$ term of neutrino propagator,
$s_{1/2}$ and $p_{1/2}$ partial waves of the electron contributes to the $0^+ \rightarrow 0^+$ transition.
However, as studied in Ref.~\cite{Tomoda1985}  and discussed in Ref.~\cite{FS},
magnetization current  $\bm{\nabla} \times \bm{\mu}$
with the  $(s_{1/2})^2$ electrons
 enhances the sensitivity of  $\eta$ term through the $V-A$ interference term.

 The enhancement mechanism of
 the $<\eta>$-term for $0^+ \rightarrow 0^+$ DBD does not work for $0^+ \rightarrow 2^+$.
In order to carry $J=2$, electrons have to be at least  $s_{1/2}$ and $p_{3/2}$ orbits.
Sensitivity of the $<\eta>$-term is similar to that of
$<\lambda>$-term  for $0^+ \rightarrow 2^+$ DBD,
while $0^+ \rightarrow 0^+$ DBD is particularly sensitive to
 the  $<\eta>$-term.

\subsection{Effective operator for $0^+ \rightarrow 2^+$ transition}

The outline of the derivation of the transition amplitude of angular momentum $J=2$ of RHC DBD
is briefly described.
Using standard partial wave expansion of the scattering wave functions of electron~\cite{HSUY},
leading order contribution of $s_{1/2}$ and $p_{3/2}$ states 
are taken as follows,
\begin{eqnarray}
  \bar{e}_{p_2,s_2}(\bm{y}) & = & \chi_{s_2}^\dagger( g_{-1}^*(p_2,y)\ , \ -f_1^*(p_2,y)(\hat{p}_2\cdot\bm{\sigma}) ), \\
  e_{p_1,s_1}^c(\bm{x}) & = &
  \left( \begin{array}{c}
    -i f_2^*(p_1,x)[
      3 (\hat{x}\cdot\hat{p}_1)(\hat{p}_1\cdot\bm{\sigma}) - (\hat{x}\cdot\bm{\sigma})]  \\
    i g_{-2}^*(p_1,x)[
      3(\hat{x}\cdot\hat{p}_1) - (\hat{x}\cdot\bm{\sigma})(\hat{p}_1\cdot\bm{\sigma})]  \\
  \end{array} \right) \chi_{s_1}^c,
\end{eqnarray}
with $\chi_{s}^c  =  i\sigma_2 \chi_s$.
$\hat{x}$ is a unit vector $\bm{x}/|\bm{x}|$.
Matrix element ${\cal M}$ is given by the sum of the each electron wave functions as
\begin{eqnarray}
  {\cal M} & = & {\cal M}^{-2-1} + {\cal M}_{2 1} + {\cal M}_{2}^{\ \ -1} + {\cal M}^{-2}_{\ \ 1}.
\end{eqnarray}
Here ${\cal M}^{ij}$ includes electron wave function $f^i_j$
defined in  Ref.~\cite{Tomoda}
where a superscript $i$ (a subscript $j$) etc. indicate that $g_i^{(-)}$ ($f_j^{(-)}$) should be taken. 
For example $f^{-2-1}=g_{-2}(p_1,R)g_{-1}(p_2,R)$, where $R$ is nuclear radius.
$f^{-2-1}$ and $f_{21}$ contribute to the phase space factor $G_1$,
while  $f^{\ \ \ -1}_{2}$  and $f^{-2}_{\ \ \ 1}$ contribute to $G_2$. 
 $G_1$ and $G_2$ are defined in
%Refs.~\cite{Tomoda1988,Tomoda2000,Fang}.
Refs.~\cite{Tomoda1988,Fang}.
${\cal M}^{-2-1}$ and ${\cal M}_{2 1}$, which are the odd element of Dirac matrix
${\cal O}$, give effective rank 2 operators of $<\lambda> M_\lambda - <\eta> M_\eta$.
${\cal M}_{2}^{\ \ -1}$ and ${\cal M}^{-2}_{\ \ 1}$ from the even element of
${\cal O}$ give $<\eta>M_\eta'$.

Matrix element  ${\cal M}^{-2-1}$ and ${\cal M}^{\ \ -1}_2$ are given as
\begin{eqnarray}
{\cal M}^{-2-1}  & = &
- i \frac{{f^{-2-1}}^*}{R}\int d\bm{x}d\bm{y} \frac{h(r)}{r^2} \chi_{s_2}^\dagger [
 <\eta> (M_{VV} + M_{AA}) + <\lambda> (M_{VV} - M_{AA} + M_{VA-}) ]
\nonumber \\
& \times &
(3  (\bm{x}\cdot\hat{p}_1) - (\bm{x}\cdot \bm{\sigma})(\hat{p}_1\cdot \bm{\sigma}))\chi_{s_1}^c, \\
{\cal M}^{\ \  -1}_{2}
& = &
 i \frac{{f^{\ \  -1}_2}^*}{R}\int d\bm{x}d\bm{y} \frac{h(r)}{r^2} \chi_{s_2}^\dagger 
<\eta> M_{VA+}
(3  (\bm{x}\cdot\hat{p}_1)(\hat{p}_1\cdot\bm{\sigma}) - (\bm{x}\cdot \bm{\sigma})
)\chi_{s_1}^c,
\end{eqnarray}
where $M_{VV},M_{AA},M_{VA\pm}$ are 
\begin{eqnarray}
  M_{VV} & = & V_0(\bm{y}) \bm{r}\cdot\bm{\sigma} V_0(\bm{x}), \\
  M_{AA} & = & (\bm{A}(\bm{y})\cdot\bm{\sigma}) (\bm{r}\cdot\bm{\sigma})
              (\bm{A}(\bm{x})\cdot\bm{\sigma}),\\
  M_{VA\pm} & = & V_0(\bm{y})(\bm{r}\cdot \bm{\sigma}) (\bm{A}(\bm{x})\cdot\bm{\sigma})
             \pm (\bm{A}(\bm{y})\cdot\bm{\sigma})(\bm{r}\cdot \bm{\sigma})V_0(\bm{x}).
\end{eqnarray}

Rewriting ${\cal M}$, the multipole operator for hadron is written
by the product of hadron and lepton irreducible tensors of rank $2$.
Here we show ${\cal M}^{-2-1}$, which include necessary operator for DBD of $\Delta_{1232}$.
The matrix element  is rewritten in terms of hadron tensor
${\cal H}$ and lepton tensor ${\cal L}$ of rank $J=2$ as
\begin{eqnarray}
{\cal M}^{-2-1}
& = & -i \frac{{f^{-2-1}}^*}{R} \frac{3\sqrt{5}}{2}[{\cal H} \otimes {\cal L}]^{(0)}, \ \ 
{\cal M}_2^{\ \ -1}
 =   i \frac{{f_2^{\ \ -1}}^*}{R} \frac{3\sqrt{5}}{2}[{\cal H}' \otimes {\cal L}']^{(0)}.
\end{eqnarray}
The lepton tensor, which is independent of nuclear coordinate is
\begin{eqnarray}
  {\cal L} &=& \chi_{s_2}^\dagger [\hat{p}_1 \otimes \bm{\sigma}]^{(2)} \chi_{s_1}^c,\ \ 
  {\cal L}' = \chi_{s_2}^\dagger( [\hat{p}_1 \otimes \hat{p}_1]^{(2)} 
  +  \sqrt{\frac{3}{2}}[[\hat{p}_1 \otimes \hat{p}_1]^{(2)} \otimes \bm{\sigma}]^{(2)}) \chi_{s_1}^c
\end{eqnarray}
and the hadron tensor is
\begin{eqnarray}
  {\cal H} = <\eta> [h_{VV} + h_{AA}] - <\lambda> [ - h_{VV} + h_{AA}  - h_{VA-}], \ \ \
  {\cal H}' = <\eta> h_{VA+}.
  \label{hadron2}
\end{eqnarray}
Here
\begin{eqnarray}
  h_{VV} & = & \int d\bm{x}d\bm{y} h(r) V_0(\bm{y})V_0(\bm{x})[\hat{r}\otimes\hat{r}]^{(2)},\\
  h_{AA} & = & \int d\bm{x}d\bm{y} h(r)(
     \frac{2}{3}[\bm{A}(\bm{y})\otimes\bm{A}(\bm{x})]^{(2)}
     - \frac{1}{3}\bm{A}(\bm{y})\cdot\bm{A}(\bm{x})[\hat{r}\otimes\hat{r}]^{(2)}
     - \sqrt{\frac{7}{3}}[[\bm{A}(\bm{y})\otimes\bm{A}(\bm{x})]^{(2)}
       \otimes [\hat{r}\otimes\hat{r}]^{(2)}]^{(2)}),
     \label{eq:aa}\\
     h_{VA-} & = & -\sqrt{\frac{3}{2}} \int d\bm{x}d\bm{y} h(r)
     {}[( V_0(\bm{y})\bm{A}(\bm{x}) +  \bm{A}(\bm{y})V_0(\bm{x}))\otimes [\hat{r}\otimes\hat{r}]^{(2)}]^{(2)},\\
     h_{VA+} & = &  \int d\bm{x}d\bm{y} h(r)\frac{r_+}{r}
     {}[( V_0(\bm{y})\bm{A}(\bm{x}) -  \bm{A}(\bm{y})V_0(\bm{x}))\otimes
       ( \frac{1}{\sqrt{2}}[\hat{r}\otimes\hat{r}_+]^{(1)} - \sqrt{\frac{3}{2}} [\hat{r}\otimes\hat{r}_+]^{(2)}) ]^{(2)}
\end{eqnarray} 
with $\bm{r}_+ = \bm{x} + \bm{y}$.

Taking the impulse approximation of the nuclear current $V_0(\bm{x})=\sum_i \tau_i^+ \delta(\bm{r}_i - \bm{x})$ and
$\bm{A}(\bm{x}) = \sum_i
g_{\rm A} \tau_i^+ \bm{\sigma}_i \delta(\bm{r}_i - \bm{x})$,
the half-life~\cite{Tomoda1988,Fang}  is given as,
%derived at first in Ref.~\cite{Tomoda1988}
\begin{eqnarray}
  (T_{1/2})^{-1} =G_1|<\lambda > M'_\lambda -  <\eta> M'_\eta|^2 +
  G_2|<\eta > M'_{\eta'}|^2,
\label{G2}
\end{eqnarray}
where
\begin{eqnarray}
  M'_\lambda=\sum_{i=1}^5C_{\lambda i}M_i,\ M'_\eta= \sum_{i=1}^5C_{\eta i}M_i,\ 
  M'_{\eta'}= \sum_{i=6}^7C'_{\eta i}M_i.  
\label{G2M}
\end{eqnarray}
We refer the coefficients $C_{xi}$ and $C_{xi}'$ to \cite{Fang,Tomoda1988}. 
 The axial vector coupling constant $g_{\rm A}^4$ is included in the phase space factor $G_i$.

\subsection{Estimation of N$\Delta$ transition in $0\nu$ DBD}
\begin{figure}[h]
  \includegraphics[width=4.5cm]{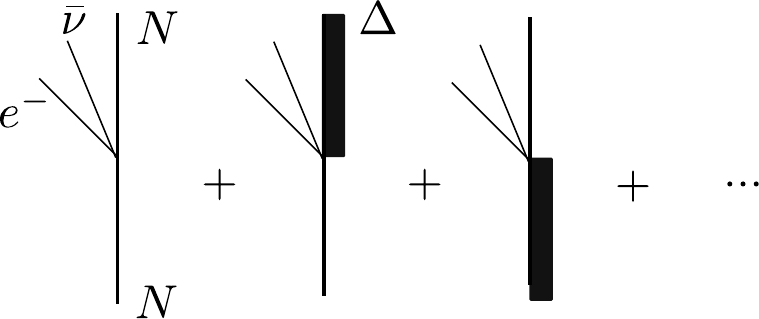}\hspace*{2cm}
  \includegraphics[width=9cm]{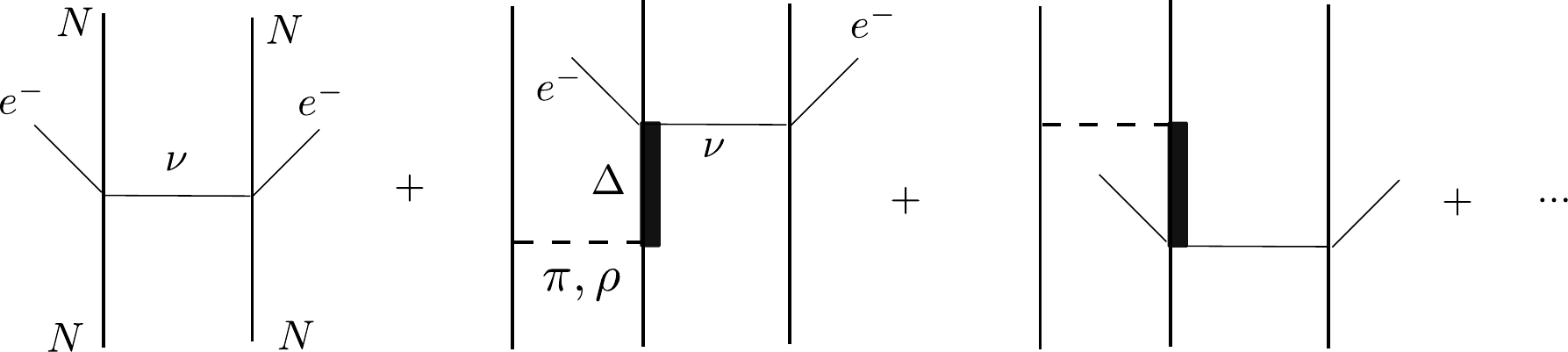}
  \caption{Semileptonic interaction of hadron. Left panel shows electron and neutrino production processes of nucleon and $N \leftrightarrow \Delta$ transition. Right panel shows 2N-mechanism of DBD.}
\label{fig:heff-dbd1}
\end{figure}

The hadron currents $J^\mu_{\rm L}$ and $J^\mu_{\rm R}$  are given in terms of current quarks. In the standard nuclear physics approach,
the nuclear state vector is described using only nucleon degrees of
freedom, instead of quarks. Therefore effective DBD interaction
Hamiltonian has to be described by using nucleon degrees of freedom.
This is achieved in two steps. At first, the semi-leptonic interaction
is truncated into hadron degrees of freedom. The interaction Hamiltonian
consists of nucleon current and
 N$\Delta$ transition current and possible
weak pion production current as shown in left panel of Fig. \ref{fig:heff-dbd1}.

\begin{figure}[h]
  \includegraphics[width=4cm]{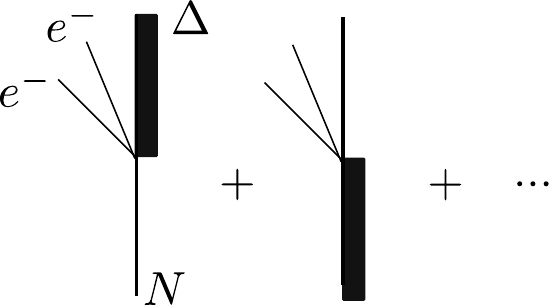}\hspace*{2cm}
  \includegraphics[width=6cm]{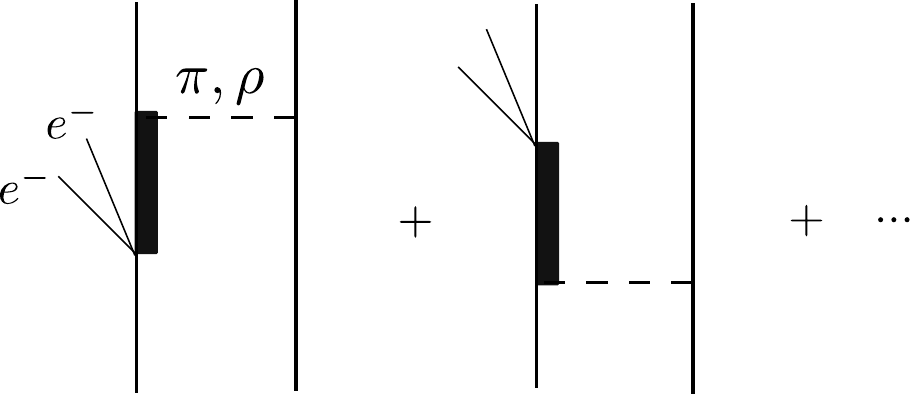}
  \caption{Diagrams of neutrino exchange within baryon. Left panel shows $0\nu$ DBD of
  $N \leftrightarrow \Delta$ transition. Right panel shows $\Delta$-mechanism of  DBD.}
\label{fig:heff-dbd2}
\end{figure}

In the second step, hadron degrees of freedom other than nucleon are eliminated 
and effective Hamiltonian to be used with many body wave function of
nucleon is obtained. The resultant effective semileptonic Hamiltonian 
is well known for beta decay, muon capture and neutrino reaction.
The nuclear weak current consists of one-nucleon impulse current (IMP)
and many-body meson exchange current (MEC).
\begin{eqnarray}
  J^\mu(\bm{x}) = J^\mu_{IMP}(\bm{x}) + J^\mu_{MEC}(\bm{x}).
  \label{eq:iamec}
\end{eqnarray}
The contribution of $\Delta$  appears as MEC.
Basic mechanism of DBD due to nuclear weak current in Eq.\bref{eq:iamec} is
the neutrino exchange between nucleons shown in the right panel of Fig. \ref{fig:heff-dbd1},
which include IMP and MEC (2N-mechanism).
The MEC of axial vector current contributes partly to
the quenching of $g_{\rm A}$ seen in beta decay~\cite{eji19a, eji22}.
Since the high momentum will be carried by exchanged neutrino in DBD
compared with beta-decay and $2\nu$ DBD,
 $g_{\rm A}^{eff}$ might depend  on the kinematics of the process. \red{Instead of $g_{\rm A}^{eff}$, the MEC in DBD has been studied in chiral effective field
theory (EFT)~\cite{menendez2011chiral,agostini2023toward}.
The MEC of EFT includes more than $\Delta$ excitation shown in the right panel
of Fig. \ref{fig:heff-dbd1}.  The MEC is estimated
with the effective one-body current obtained by applying the 'normal ordering'
and is shown to give an important mechanism of the reduction of the Gamow-Teller transition.
Here, we concentrate on the $\Delta$ mechanism which emerges in DBD 
as shown  in left panel of Fig. \ref{fig:heff-dbd2}~\cite{doi85,Tomoda}.}
Neutrino is exchanged between quarks within single hadron.
This mechanism generates new MEC as shown in the
right panel of  Fig. \ref{fig:heff-dbd2}.
Among the excited nucleon states, resonance must be iso-spin
$3/2$ $\Delta$ resonance.
$\Delta_{33}(1232)$ is expected to give the largest contribution.
Then  N$\Delta$ DBD contributes
for $0^+ \rightarrow 2^+$ transition($\Delta$-mechanism) but not for $0^+ \rightarrow 0^+$ transition.

Within the non-relativistic constituent quark model with SU(6) symmetry,
the matrix element of $h_{AA}$ for $n \rightarrow \Delta^{++}$ transition is
given as
\begin{eqnarray}
  {\cal M}_\Delta  = <\Delta^{++}||h_{AA}||n> =
  < \Delta^{++}||\sum_{i,j=1}^3 \frac{2}{3}[\bm{\sigma}_i \otimes\bm{\sigma}_j]^{(2)}h(r_{ij})||n> = - \frac{8\sqrt{10}}{3}<h>,
\end{eqnarray}
where the first term of $h_{AA}$ in Eq. \bref{eq:aa} in terms of quark operator contributes.
  The $\Delta$-mechanism contributes to both $\eta$ and $\lambda$ terms.

The $\Delta$-mechanism is estimated in Ref.~\cite{doi85}.
They give a large contribution of the $\Delta$ mechanism.
It is estimated by using 
 the probability of the $\Delta$ in nuclear wave function
 $P(\Delta)$ and the transition matrix element of the $A-1$ nucleons $<\Phi_F|\Phi_I>$.
$P(\Delta$) is estimated as $P(\Delta) \sim 0.01$ following Ref. \cite{Primakoff}
based on the probability of iso-spin 1/2 $N^*$ resonances instead of $\Delta$,
suggested by the analysis of magnetic moment and $(p,d)$ reaction at that time. 
The transition matrix element is assumed to be $<\Phi_F|\Phi_I> \sim \sqrt{0.1}$.
Using those numbers, $\Delta$-mechanism is  estimated as,
\begin{eqnarray}
  \frac{<2^+||h_{AA}||0^+>}{|{\cal M}_\Delta|}  & = &  \sqrt{P(\Delta)}<\Phi_F|\Phi_I>
  \sim 3.2 \times 10^{-2}. \label{eq:doi}
\end{eqnarray}

The $\Delta$-mechanism is studied in the microscopic model
for the DBD of $^{76}$Ge in Ref.~\cite{Tomoda1988}.
The $\Delta$ component in nuclei is evaluated perturbatively by using the 
transition potential $V_{NN \leftrightarrow N\Delta}$,
\begin{eqnarray}
  <2^+||h_{AA}||0^+> & = & <F||V_{NN-N\Delta}\frac{1}{E_f - H}<\Delta|h_{AA}|N>
  +<N|h_{AA}|\Delta>\frac{1}{E_i - H}V_{N\Delta-NN} ||I>.
\end{eqnarray}
The NN-N$\Delta$ transition potential is given by
the pion and rho meson exchange model.
The energy denominator is approximated by the
 N$\Delta$ mass difference $E_{i/f} - H \sim m_N - m_\Delta$.  The $\Delta$-mechanism is evaluated by
the meson-exchange current in Fig.~\ref{fig:heff-dbd2}
with 'Hartree-Fock-Bogoliubov type' nuclear wave function.
The $\Delta$-mechanism of the microscopic model turn out to be about $1/60$ of
the estimation of (\ref{eq:doi}).
\red{
  The contribution of the $\Delta$-mechanism 
  is not very large compared with the usual 2N-mechanism~\cite{Vergados:1988xp}.
}

The dominant contribution of the $\Delta$-mechanism is due to the matrix
element of the following operator:
\begin{eqnarray}
\sum_{m,n=1}^A \tau_m^+\tau_n^+ \bm{\sigma}_m\cdot\bm{\sigma}_n [\hat{r}_{mn}\otimes\hat{r}_{mn}]^{(2)}.
\end{eqnarray}
The ratio of the NME of operators which have 
the same spin and angular momentum structure for both the 2N- and $\Delta$-mechanism
gives us less model dependent estimation of the $\Delta$-mechanism.
Using the notation of Ref.~\cite{Tomoda1988}, $M_1$ is for the 2N-mechanism
and $M_9$ and $M_{12}$ are for the $\Delta$-mechanism.
Apart from coupling constants,  difference among the matrix element $M_i$ is 
only radial dependence of the operators,
which is $h(r)$ for 2N-mechanism while that of the $\Delta$-mechanism with one-pinon-exchange is
$Y_2(m_\pi r) = e^{-m_\pi r}(1 + 3/(m_\pi r) + 3/(m_\pi r)^2)/r$.
Using the numbers of Ref.~\cite{Tomoda1988}, the ratio is calculated as
\begin{eqnarray}
  r & = & \frac{C_{\lambda 1} M_1 }{ c_9 M_9  + c_{12} M_{12}} \sim 0.12.
\end{eqnarray}
Assuming the A-dependence of the radial overlap integral is weak,
the $\Delta$-mechanism for this particular type of matrix element is estimated as about $10\%$ of that of
the 2N-mechanism using
$g_{\rm A}=1.27$, while the $\Delta$-mechanism can be $20\%$ using
quenched $g_{\rm A}^{eff} \approx 0.7$~\cite{eji22}.

%%%%%%%%%%%%%%%%%%%%%%%%%%%%%%%%%%%%%%%%%%%%%%%%%%%%%%%%%%%%%%%%%%%%%%%%%%
\red{\section{RHCs to be studied by ton-scale DBD experiments}}
We discuss first  DBD rates for the individual RHC processes, i.e. the $<\lambda>$ and $<\eta>$ dominant cases, to see how the rates depend on $<\lambda>$ and $<\eta>$. 
The rate $\Gamma_k$ with $k$ being $\lambda$ and $\eta$ for the
 0$^+\rightarrow $ 0$^+$ ground-state ($J$=0)  and the 0$^+\rightarrow$ 2$^+$ excited state ($J$=2) transitions are given as in eqs. (11) and (49) and in \cite{doi85,eji19,eji20,suh98} as \red{
\begin{equation}
\Gamma_{k}^{(J)}=C_{kk}^{(J)} (<k>)^2, 
\end{equation}
where $C_{kk}^{(J)}$  is the sensitivity to the $k$-mode DBD decay. The sensitivity is a kind of the amplification factor to make the small $<k>$ beyond SM visible experimentally.  Actually, $C_{kk}^{(J)}$ is given by $\Sigma_iG_{kk}^{(J)}(i) [M_{kk}^{(J)}(i)]^2$ with $G_{kk}^{(J)}(i)$ and $ [M_{kk}^{(J)}(i)]$ being  the relevant  $\beta\beta$ phase space factors and the relevant NMEs. }They correspond, respectively, to the atomic and nuclear amplification factors. The axial-vector weak coupling factor of
$g_{\rm A}^{\ 4}$, where $g_{\rm A}$ is
 the axial-vector coupling in units of the vector coupling of $g_{\rm V}$ for a free nucleon, is conventionally included in $G_{kk}^{(J)}$. 

 \red {The phase space factor $G_{kk}^{(J)}(i)$ depends on the DBD $Q$ value and the atomic number $Z$ of the DBD nucleus, and is rather well calculated in past and also in recent works \cite{doi85,kot12,sto19,Fang}}.
The DBD NME is known to be rather sensitive to the nuclear models and the nuclear parameters used for the model calculation \cite{eji19}. The NMEs involved in RHCs include various kinds of  spin ($\sigma$) isospin ($\tau$) components, which depend much on the nuclear and non-nuclear $\sigma \tau$ correlations. Then the $\sigma\tau$ NMEs are known to be quenched much due to the $\sigma\tau$ correlations. Accordingly, the effective axial-vector coupling of $g_{\rm A}^{\rm eff}$ is used for the model NME to incorporate such quenching effects as the nuclear medium effects and the non-nucleonic $\sigma\tau$ correlation effects that are not included explicitly in the model NME. 

Recently $g_{\rm A}^{\rm eff}/g_{\rm A}$ is shown to be around 0.55 by analysing the single $\beta$
decay rates and the summed $\sigma\tau$ strength in medium-heavy nuclei \cite{eji22,eji25}.

\red {The DBD rates for the RHCs are evaluated for typical DBD nuclei of $^{76}$Ge, $^{82}$Se,  $^{100}$Mo, $^{130}$Te and $^{136}$Xe. These nuclei are of current interest since good energy-resolution detectors and large amounts of the DBD isotopes are available and thus future ton-scale experiments are under progress by using these isotopes \cite {ago23,ume24}. }

The $\Gamma_{k}^{(0)}$ for the ground state is evaluated by using the  NMEs based on the QRPA (Quasi-Particle Random Phase Approximation) model \cite{mut87,Tomoda} and the shell model \cite {fan24}, and $\Gamma_{k}^{(2)}$ for the excited 2$^+$ state by using the QRPA NMEs \cite{fan24}. Note that these models take explicitly into counts the $\sigma\tau$ and other correlations. Then the quenching coefficient of  $g_{\rm A}^{\rm eff}$/$g_{\rm A}\approx$ 0.55 \cite{eji22,eji25} may be used to incorporate the nuclear medium and non-nucleonic $\sigma\tau$ correlations that are not included in their models.

\begin{figure}[hb]
%\hspace{0cm}
  %\vspace{-0.3cm}
\includegraphics[width=0.7\textwidth]{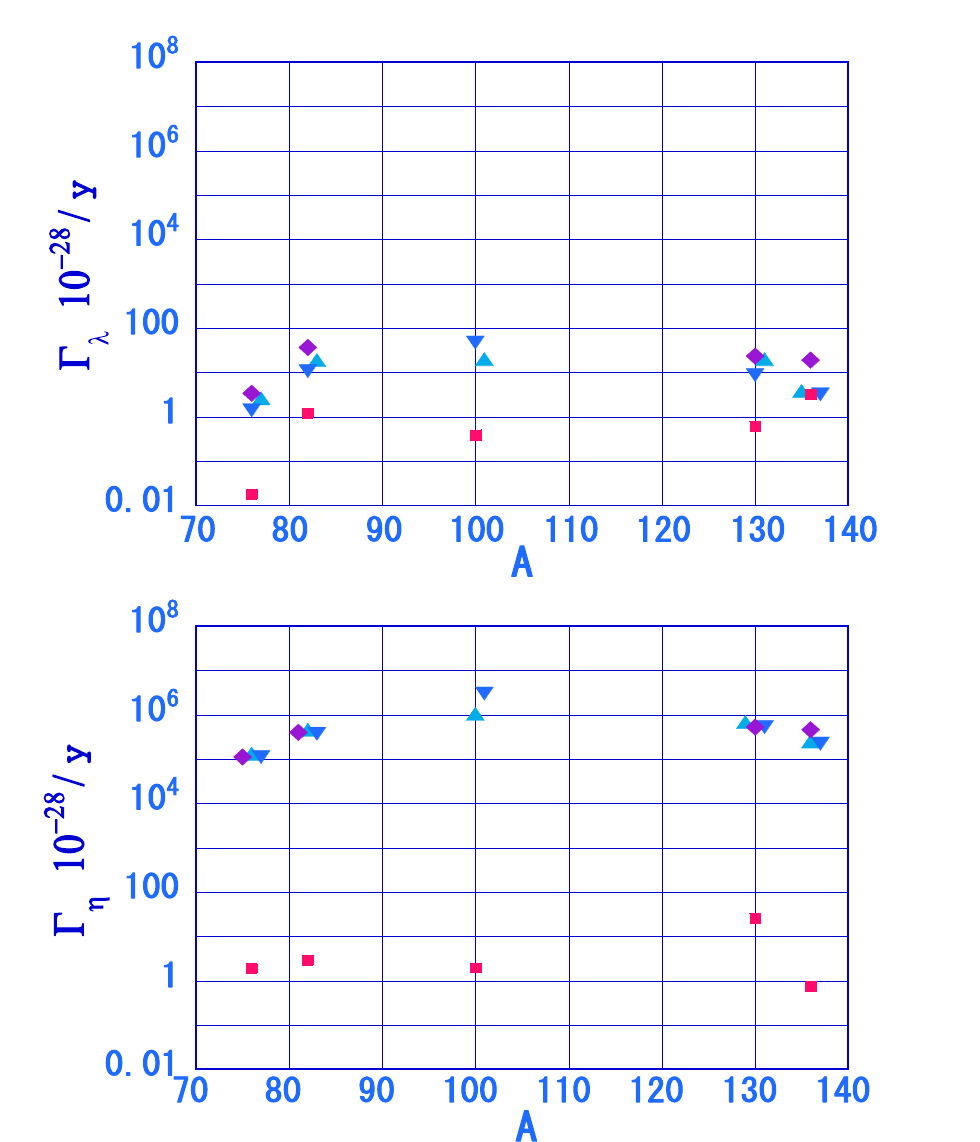}
%\hspace{7cm}
%\vspace{0.3cm}
\caption{RHC rates  ($\Gamma_k=(C_{kk}$) in units of 10$^{-28}$ per year for RHC
%  $<\lambda>$=1 and $<\eta>$=1
  $k$=1 
  in units of 10$^{-7}$
for typical DBD nuclei of $^{76}$Ge, $^{82}$Se,  $^{100}$Mo, $^{130}$Te, $^{136}$Xe of current interest. 
Top: $k=<\lambda>$. Bottom: $k=<\eta>$. NMEs used are  the QRPA model NMEs in \cite{mut87} (light blue triangles), the QRPA model NMEs in \cite{Tomoda} (blue inverse triangle) and the shell model NMEs in \cite{fan24} (violet diamond) for the ground-state (0$^+$) transitions and the QRPA NMEs in \cite{Fang} for the excited 2$^+$ transitions (red square).  Some $\Gamma _k$ points are shifted by the mass number 1 or 2 to avoid overlap with the others.   
\label{figure:fig1}} 
\end{figure}

\red {The obtained RHC rates are shown in Fig. 3.  
 $\Gamma_{\lambda}^{(0)}$ in units of 10$^{-28}$/y for $\lambda$=1 in units of 10$^{-7}$ is around 10,  while $\Gamma_{\eta}^{(0)}$ in the same units  are of the order of 10$^6$, reflecting the much larger phase space factor of $G_{09}$ for the recoil term
 involved in the $\eta$ current for the ground-state decay, as discussed in section 3. On the other hand, $\Gamma_{\lambda}^{(2)}$ for $\lambda$=1 in the same units is scattered around 1, and $\Gamma_{\eta}^{(2)}$ scatter around 5.  The ratio of $
\Gamma_{\lambda}^{(2)}$/$\Gamma_{\lambda}^{(0)}\approx$0.1 reflects the ratio for their phase space factors due to the smaller $Q$ values for the excited states. We note that the RHC rates for given $\lambda$ and $\eta$ are smaller for $^{76}$Ge by an order of magnitude than those for others. The smaller rates are due to the smaller phase space factors because of the smaller $Q$ values for $^{76}$Ge. } 

The $\nu$-mass of $<m>$ and the RHCs of $<\lambda>$ and $<\eta>$  to be studied are beyond the standard model, and thus are considered to be very small if they are non-zero.  No signals so far in $0\nu$ DBD experiments with typical medium-heavy DBD nuclei suggest that the $0\nu$ DBD rate for the ground-state transition is of the order of $\Gamma_{k}$=10$^{2}$/y in units of 10$^{-28}$ or less , i.e. $T_{1/2} \ge10^{26}$ y. This limit excludes the quasi-degenerate effective $\nu$-mass of  $<m> \ge$100 meV, depending on the NME. 
Thus next generation ton-scale DBD experiments are under-progress or planned to search for the DBDs in the region of the IH mass region of the effective $\nu$-mass around $<m>\approx 0.6 \times 10^{-7}$ in units of $m_e$. This corresponds to the decay rate of $\Gamma_{m}$=1 /y in units of 10$^{-28}$ or less , i.e. $T_{1/2} \ge10^{28}$ y. 

Now we discuss how the ton-scale detectors are used to access the new RHC region of interest.  So, let us evaluate the minimum (i.e. the lower limit of) RHCs $\lambda_{min}$ and $\eta_{min}$ to be studied by the ton-scale experiments. They are obtained by requiring that the number of the DBD signals should exceed the fluctuation of the background signals, and thus depends on the detector specifications (number of the isotopes, background rate, etc) and the run time.  The minimum RHC rates $\Gamma_{min}$ in units of 10$^{-28}$ to be studied by a DBD experiment  is given as \cite{eji19}
\begin{equation}
\Gamma_{min} (d) =0.24 (A/100) d^2, ~~~d=1.3 \epsilon^{-1/2} (B/NT)^{1/4},
\end{equation}
where $A$ is the mass number of the DBD isotope, $d$ is the sensitivity of the experiment with $N$ ton isotopes, $B$ per t y (ton year) backgrounds and $T$ y (year) run time.  The minimum RHC to be studied is given by $k_{min}=(\Gamma_{d}/C_{kk})^{1/2}$.   Note that $k_{min}$ depends linearly on the NMEs in $C_{kk}$ and the $d$, and thus very weakly on $B/N$ to the power of 1/4.  It requires 100 times more $N$ and 100 times less $B$ to reduce the minimum RHC by a factor 10. 

The minimum RHC rate by a typical ton scale DBD experiment for the DBD isotope with $A$=100 and $d$=1 with $N$=1 t, $B$=1 per t y, $\epsilon$=0.7, and $T$=5 y is $\Gamma(J)$=2.4 in units of 10$^{-28}$ for both $<\lambda>$ and $<\eta>$ in case of typical calorimetric detectors to measure the sum of the two $\beta$ rays. 

We first consider DBD experiments with $d$=1 and  $\Gamma^{(J)}$=2.4 for both the ground-state ($J$=0) and excited-state ($J=2$) transitions by
 such calorimetric detectors as to measure the total energy of the $\beta\beta$-rays and the
 $\gamma$-ray. Note that the total energy for the excited-state transition is the same as that for the
 ground state one. Then, the $\lambda_{min}$ and $\eta_{min}$ to be studied are shown in Fig. 5.
Here we used the rates of $\Gamma_{\lambda}^{J}$=10 and 1 for $J$=0 and 2 in units of 10$^{-28}$ for $\lambda$ =1 in units of 10$^{-7}$, and  $\Gamma_{\eta}^{J}$=10$^6$ and 5 for $J$=0 and 2 in the same units (see Fig. 3). 

\begin{figure}[ht]
%\hspace{-0.cm}
\includegraphics[width=0.8\textwidth]{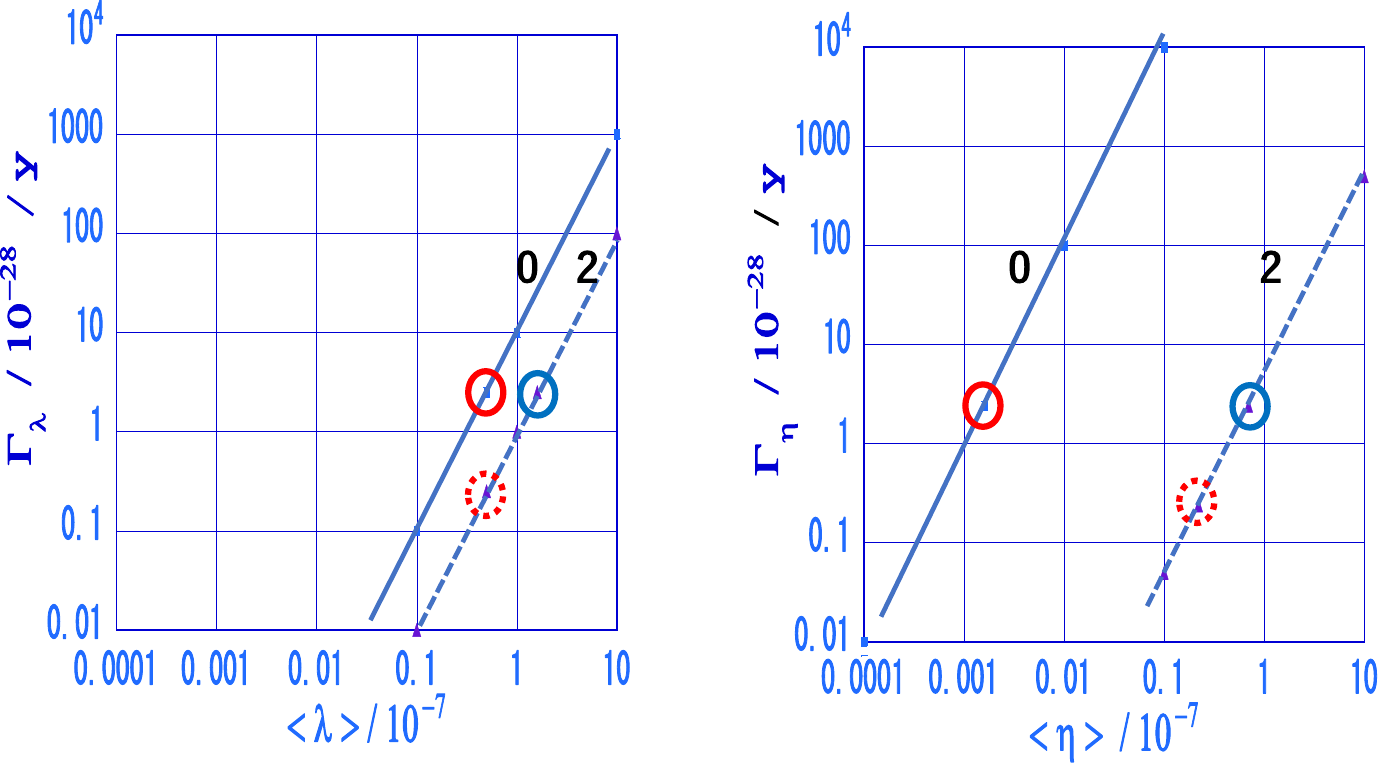}
%\vspace{-2.0cm}
\caption{The RHC rate $\Gamma_{\rm k}$ against the RHC $<\rm k>$ for the ground-state (solid line) and the excited-state (dotted line) transitions. Left panel: k=$\lambda$.  Right panel: k=$\eta$. Solid circles: Minimum RHCs to be studied by a typical ton-scale experiment of $d$=1 without $\gamma$-coincidence. Dotted circles: Those by a typical ton-scale experiment for $d$=0.01 with $\beta\beta-\gamma $ coincidence. See text.  
\label{figure:fig5}} 
\vspace{-0.cm}
\end{figure}
\red {The minimum RHCs to be studied with a typical ton-scale experiment with $d$=1 is $<\lambda_{min}>\approx$0.5 for the ground state and $<\lambda_{min}>\approx$1.5 for the excited state, while $<\eta_{min}>\approx 1.5 \times10^{-3}$ for the ground state and $<\eta_{min}>\approx$0.7 for the excited state, all in units of 10$^{-7}$, as shown in the left and right panels of Fig.4. These values reflect the square roots of the phase space factors. $<\lambda_{min}>$ for the excited state is less stringent by a factor 3 than that for the ground state, but is used to confirm the $<\lambda>$ if observed in the ground state transition, while $<\eta_{min}>$ for the excited state is too large to be observed since $<\eta>$ itself is of the order of 10$^{-9}$ or less. The DBD rate is proportional to the phase space factor $G$, which is roughly proportional to $Q^5$, and the $Q$ value for the excited state is about 2/3 of the ground state $Q$,  and thus the DBD rate for the excited state is an order of magnitude smaller than that for the ground state if the NMEs are similar for both. Accordingly, the DBD to the excited state is so far considered to be of less discovery potential for exploring the RHC. } 

\red {The present work shows that a  better-sensitivity (smaller $d$) measurement for the excited state is possible by measuring the $\gamma$ ray from the 2$^+$ state in coincidence with the $\beta\beta $ rays to the 2$^+$ state. Noting that the $\beta$ flight-length is of the order of mm, while the $\gamma$ one is of the order of cm,  multi-cell (multi-segmented) detectors is effective to separate the excited state DBD followed by the $\gamma $ ray from the ground-state DBD since the 
$\beta-\beta$ starting at the cell C$_1$ deposit their energies on the same cell C$_1$, while the $\gamma$ may pass through the cell C$_1$ and deposits the energy in the cells C$_i$ around the cell
 C$_1$. Then the ground state DBD is selected by setting the energy window at $E=Q$ for the cell
 C$_1$ with the multiplicity=1, while the excited-state DBD by the energy window at $E=Q'$ for the cell C$_1$ and the energy window at $E=E_{\gamma}$ for the summed signals from the cells C$_i$ around. 
Thereby the backgrounds are much reduced, depending on the energy window for the $\gamma$.}
The backgrounds from the 2$\nu$ DBD decay to the excited state is also much reduced by a factor around $(Q'/Q)^{11}\approx$0.1 than that for the ground-state transition. Back grounds from solar neutrinos are also reduced much \cite{eji14,eji17}. 

Actually, the background reduction rate depends much on the energy window, namely the energy resolution of the $\gamma $ measurement. It is evaluated to be of the order of  $B'/B\approx$0.01 in case of bolometers with the energy resolution of around 10 keV. Then the detector sensitivity $d$ gets smaller by a factor around $(0.01)^{1/4}\approx$0.3 by requiring coincidence with the $\gamma$ ray, resulting in the smaller RHC rate and the smaller RHC by factors around 0.1 and 0.3, respectively, as shown in the doted circles. Then one can search for the similar $<\lambda>$ region around 0.5 in units of 10$^{-7}$ by both the ground-state  and excited-state decays. 

\red {The region of the minimum RHCs of $<\lambda_{min}>$ and $<\eta_{min}>$ to be studied by the typical ton-scale experiment is shown in Fig.5. Here we use the sensitivity of $d=1$ for the ground-state transition and $d$=0.3 for the excited-state transition by the $\beta-\gamma$ coincidence. It is out of, but is close to  the theoretically suggested region of 1 $\ge$ tan$ \beta$ $\ge$165 with tan $\beta$ being around $<\lambda>/<\eta>$. }

\begin{figure}[ht]
%\hspace{-0.cm}
\includegraphics[width=0.5\textwidth]{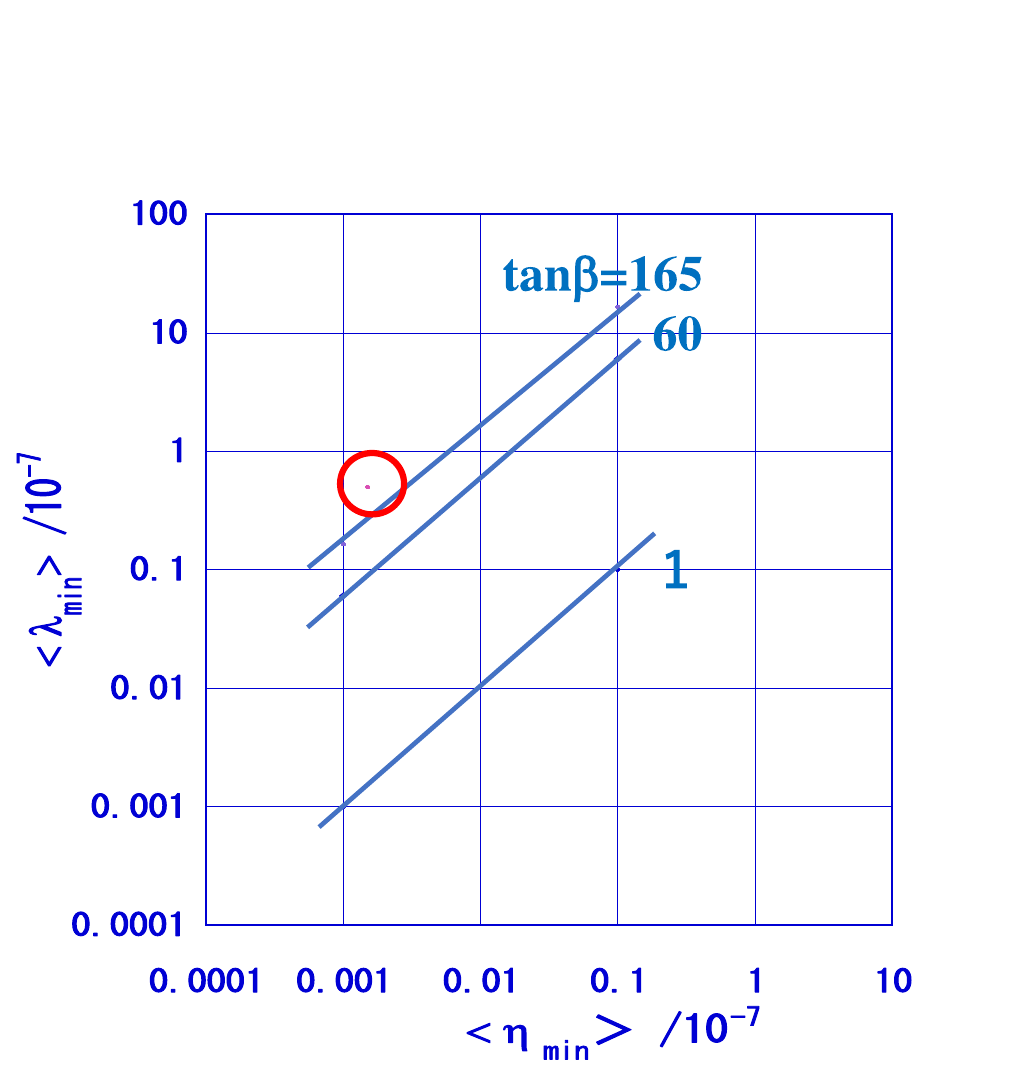}
%\vspace{-1.cm}
\caption{Pink circle : the region around minimum RHCs of $<\lambda_{min}>$ and $<\eta_{min}>$ to be studied by the typical ton-scale experiment with the sensitivity of $d=1$ and $d$=0.3 for the ground- and excited-state decays. The solid lines show the lower bound of 1 and the upper bounds of 60  and 165 for tan$\beta$, which is related to the order of the $<\lambda>$ to $<\eta>$ ratio.  See text.  
\label{figure:fig6}} 
\vspace{0cm}
\end{figure}

\red {The regions of the  $<\lambda>$ and $<\eta>$ to be studied by the ton-scale DBD experiments of both the ground-state and excited-state decays  are summarized as follows.}

\red{A: The $\beta\beta-\gamma$ coincidence measurement makes it possible to get the detector sensitivities around $d$=1 for the ground-state decay with no $\gamma$-coincidence  and $d\approx$0.3 for the excited-state decay with the $\gamma$-coincidence. } 

\red {B: The $<\eta>$ region above around $1.5\times 10^{-3}$ in units of $10^{-7}$ is well studied by searching for the $\beta\beta$ signals in the ground-state channel.  No signals from the  $<\eta>$ in the excited-state channel. }

\red {C: The $\lambda$ region above around 0.5 in units of 10$^{-7}$ is exclusively studied by searching for the excited-state decay.
%  Experimantal observation
 If one observes 
 non-zero signals in the excited-state window and no signal in the ground state channel, 
 the ratio of $<\lambda>/<\eta>$ is constrained to be of the order of 300 or larger, and
 the  SUSY is disfavored.
% thus disfavor the SUSY.
 Thus it provides experimentally the favored region of tan$\beta$ and the models behind it. }

\red {D: In case of non-zero signals in both the ground- and excited-state windows, the signals in the ground state window may be due to both the $\lambda$ and $\eta$ currents.  Here the $\lambda$ contribution in the ground-state window may in principle be evaluated from the rate in the excited state channel if the RHC NMEs for both the ground- and excited-state decays are well established. }

\red {E: In other words, the ratio $R(0/2)$ of the DBD rates for the ground $0^+$ and excited $2^+$ state decays is used to study exclusively the individual $\lambda$ and $\eta$ currents, $R(0/2)\gg1$ leading to the $\eta$ dominance and $R(0/2)\ll$1  to the $\lambda$ dominance.  If we know well the relevant NMEs, the ratio is used to get the individual $\eta$ and $\lambda$. }

So far we have discussed the RHC rate using mainly the NMEs based on QRPA models with nuclear $\sigma\tau$ correlations and the quenched $g_{\rm A}^{\rm eff}$ with non-nuclear $\sigma\tau$ correlations. In fact there are several models to evaluate the DBD NMEs, and the evaluated NMEs scatter over an order of magnitude, depending on the models \cite{eji19}.  Actually the recent IBM (Interacting Boson Model) NMEs \cite {fer23} for the 2$^+$ RHCs are smaller by an order of magnitude than the QRPA ones. These smaller NMEs lead to $<\lambda_{min}>$ around 5 in units of 10$^{-7}$ for the excited-state transition. Then one may have no DBD signal in the 2$^+$ window in case of $<\lambda>\approx 10^{-7}$.

\red {It is noted that the transition amplitude (square root of the transition rate) of the NH $\nu$-mass term of $<m_{\nu}>/m_e\approx 0.05$ is smaller by an order of magnitude than the amplitudes of the RHC $<\lambda>\approx 0.5 $ and
 $<\eta>\approx1.5 \times 10^{-3}$, all in units of 10$^{-7}$,  to be studied by the ton-scale detector for the ground state transition. It is smaller by 3 orders of magnitude than that for the RHC  $<\lambda>\approx 0.5 $ to be studied by the ton-scale detector for the excited state transition \cite{Tomoda,Tomoda2000}. Thus almost no contributions from the $\nu$-mass term in the ton-scale experiment}.  

\red {The possible regions of RHCs to be studied were discussed for simplicity for diagonal elements of  
$\Gamma_{k}^{(J)}=C_{kk}^{(J)} (<k>)^2$ with $k$ being $\lambda$ and $\eta$, but the non-diagonal term of $C_{\lambda,\eta}(<\lambda><\eta>$) plays some role, depending on the NMEs, in case of the ground state transition with non-zero $\lambda$ and $\eta$ currents.}

\section{Concluding remarks}

\red {1. The present paper discusses new theoretical and experimental aspects of  the RHCs of $<\lambda>$ and $<\eta>$ to be studied by the next-generation ton-scale DBD detectors. They are under progress to explore DBDs in the region of $T_{1/2}\approx 10^{-28}$ y to search for the effective $\nu$-mass in the IH region. The $\nu$-mass spectrum, however, may likely be NH, and then no signal from the mass term.  Then the detectors are used to explore the new regions of the RHCs.}

2. \red {On the basis of the left-right symmetric model, further theoretical discussions on the regions of the RHCs are made and the order of the ratio of $<\lambda>$/$<\eta>$ is shown to be given by tan$\beta$, which is confined to be 1$\le$ tan$\beta$ $\le$60, 165, depending on the models of BSM.}

3. The NMEs involved in the RHCs have been explicitly given 
for both the $0^+$ ground-state and  $2^+$ excited-state decays.
The enhancement mechanism of the $<\eta>$-term in the $0^+$ transition has been clearly shown.
\red {The $\Delta$-mechanism with a $\nu$ exchange between quarks in $\Delta$, which
appears in the $2^+$ decay, is shown to be about 20$\%$ of the 2N-mechanism for the leading
axial vector current NME of $\bm{\sigma}_m\cdot\bm{\sigma_n} Y_2$
in case of the quenched $g_{\rm A}^{eff}\approx 0.55 g_{\rm A}$ for the 2N.}

\red {4. The new  regions of RHCs are shown to be studied exclusively by measuring both the ground-state $\beta\beta$ decay followed  by no $\gamma$ ray and the excited-state $\beta\beta$ decay followed by the $\gamma $ ray.  Here multi-cell (segmented) detectors like bolometers are used to measure separately the $\beta \beta$ and the $\gamma$.  Then the $<\lambda>$ region above 0.5 in units of
 10$^{-7}$ and the $<\eta>$ region above 1.5 10$^{-3}$ in units of 10$^{-7}$ are studied, and   non-zero values for or limits on $<\lambda>$, $<\eta>$ and  $<\lambda>$/$<\eta>$ give great impact on the BSM physics.}

\red {5. The RHCs to be studied are evaluated by using mainly QRPA NMEs with nucleonic $\sigma\tau$ correlations and the quenched coupling of $g_{\rm A}^{\rm eff}$ to
incorporate the non-nucleonic $\sigma\tau$ correlations and others. Actually the NMEs depend on the models and parameters used for the model as in case of the NMEs for the $\nu$-mass mode. Needless to say, accurate theoretical and experimental evaluations for the $\lambda$ and $\eta$ NMEs are indispensable for quantitative studies of them.}

\red{6. It is encouraged to study RHCs on several nuclei to confirm the RHCs, if they are observed, in views of the uncertainties of the NMEs and the fluctuations of the backgrounds at the region of interest. They would help to identify the RHCs involved if the individual RHC NMEs would be very accurately evaluated and their values would be very different among the nuclei. }

\red {7. Alternatively, experimental studies of the 2$^+$  DBDs in more than two nuclei provide a unique opportunity to check if the NMEs are right or wrong since $<\lambda>$, being the same among the nuclei, is only one BSM term involved in the L-R symmetry model. }

\red {8. The DBD detectors discussed in the present work are calorimetric detectors with high discovery potential. High sensitivity ton-scale tracking detectors are of great interest to  measure the $\beta$-$\beta$ energy correlation to identify the individual $\lambda$ and $\eta$ currents \cite{doi85,eji05,ver12,ste15,Tomoda,Tomoda2000}. } 

\section{Acknowledgments}
\red{We are grateful to K.S. Babu and N. Okada for the comments on Eq.\bref{tanbeta}}. This work was partly supported by JSPS KAKENHI Grants No. 23K22508 (T.F.).

\end{document}